\documentclass[final,number,5p,twocolumn]{elsarticle}
\usepackage{graphicx}
\newcommand{\bvec}[1]{\mbox{\boldmath $#1$}}

\journal{Astroparticle Physics}

\begin{document}

\begin{frontmatter}

\title{Relativistic corrections to the Kompaneets equation}

\author{Satoshi Nozawa}
\ead{snozawa@josai.ac.jp}
\address{Josai Junior College, 1-1 Keyakidai, Sakado-shi, Saitama, 350-0295, Japan}
\author{Yasuharu Kohyama}
\address{Advanced Simulation Technology of Mechanics Co. Ltd, 2-3-13 Minami, Wako-shi, Saitama, 351-0104, Japan}

\begin{abstract}
  We study the Sunyaev$-$Zeldovich effect for clusters of galaxies.  We explore the relativistic corrections to the Kompaneets equation in terms of two different expansion approximation schemes, namely, the Fokker-Planck expansion approximation and delta function expansion approximation.  We show that two expansion approximation formalisms are equivalent under the Thomson approximation, which is extremely good approximation for the CMB photon energies.  This will clarify the situation for existing theoretical methods to analyse observation data.

\end{abstract}


\end{frontmatter}

%% main text

\section{Introduction}

  The Sunyaev$-$Zeldovich (SZ) effect \cite{zeld69,suny80}, which arises from the Compton scattering of the cosmic microwave background (CMB) photons by hot electrons in clusters of galaxies (CG), provides a useful method for studies of cosmology.  For the reviews, for example, see Refs.~\cite{birk99} and \cite{carl02}.  The original SZ formula has been derived from the Kompaneets equation \cite{komp56} in the non-relativistic approximation.  However, X-ray observations, for example, by Allen et al. \cite{alle02} have revealed the existence of high-temperature CG such as $k_{B} T_{e} \simeq $20keV.  For such high-temperature CG, the relativistic corrections will become important.

  The theoretical studies on the relativistic corrections have been done by several groups.  Wright \cite{wrig79} and Rephaeli \cite{reph95} have done the pioneering work to the SZ effect for the CG.  Challinor \& Lasenby \cite{chal98} and Itoh, Kohyama \& Nozawa \cite{itoh98} have adopted a relativistically covariant formalism to describe the Compton scattering process and have obtained higher order relativistic corrections to the thermal SZ effect in the form of the Fokker$-$Planck expansion approximation.  Nozawa, Itoh \& Kohyama \cite{noza98} have extended their method to the case where the CG is moving with a peculiar velocity $\bvec{\beta}_{\rm{c}}$ with respect to the CMB frame and have obtained the relativistic corrections to the kinematical SZ effect.  Itoh, Nozawa \& Kohyama \cite{itoh00} have also applied the covariant formalism to the polarization SZ effect \cite{suny81}.  The effect of the motion of the observer was also studied, for example, by Refs.~\cite{chlu05} and \cite{noza05}.

  On the other hand, Brown \& Preston \cite{brow12} studied the leading order relativistic corrections to the Kompaneets equation by calculating the Boltzmann equation in the delta function expansion approximation.  Although their expansion scheme is quite different from the previous schemes \cite{chal98,itoh98}, their leading order corrections on the SZ effect reproduced the previous results \cite{chal98,itoh98}.

  In this paper, we study the SZ effect for clusters of galaxies.  We explore the relativistic corrections to the Kompaneets equation in terms of two different expansion approximation schemes, namely, the Fokker-Planck expansion approximation \cite{noza09} and delta function expansion approximation \cite{brow12}.  We show that two expansion approximation formalisms are equivalent under the Thomson approximation.  This will clarify the situation for existing theoretical methods to analyse observation data.

  This paper is organized as follows.  In Section 2, we study the Boltzmann equation in the Fokker-Planck expansion approximation under the Thomson approximation \cite{noza09}.  We derive analytic expressions for the Boltzmann equation which is identical to Ref.~\cite{noza14}.  In Section 3, we investigate the Boltzmann equation in the delta function expansion approximation \cite{brow12}.  Starting from the expression in the delta function expansion approximation, we derive the same expression as the Fokker-Planck expansion approximation.  Finally, concluding remarks are given in Section 4.

\section{Fokker-Planck expansion formalism}

\subsection{Boltzmann equation}

  Let us consider that both the CG and observer are fixed to the CMB frame.  Then, the Boltzmann equation for the electron-photon scattering in the CMB frame is written by:
\begin{eqnarray}
\frac{{\rm{d}} n(\omega)}{{\rm{d}} t} = - 2 \int {\rm{d}}^{3}p \, {\rm{d}}^{3}p^{\prime} \, d^{3}k^{\prime} \,  W  \, \left[ n(\omega) \left\{1 + n(\omega^{\prime}) \right\}  \right. \nonumber  \\
&& \hspace{-60mm}
\left. \times f(E) - n(\omega^{\prime}) \left\{1 + n(\omega) \right\} f(E^{\prime}) \right] \, ,
\label{eq2-1-1}  \\
&&\hspace{-82mm}
W = \frac{\delta^{4}(p+k-p^{\prime}-k^{\prime}) \, \alpha^{2} \, \bar{X}}{(2\pi)^{3} 2 \omega \omega^{\prime} E E^{\prime}}  \, ,
\label{eq2-1-2}
\end{eqnarray}
where $k=(\omega,\bvec{k})$ and $k^{\prime}=(\omega^{\prime},\bvec{k}^{\prime})$ are the initial and final CMB photon momenta, respectively, and $p=(E,\bvec{p})$ and $p^{\prime}=(E^{\prime},\bvec{p}^{\, \prime})$ are the momenta for electrons.  In Eq.~(\ref{eq2-1-1}), $n(\omega)$ and $f(E)$ denote the distribution functions for the CMB photons and electrons in the CG, respectively.  In Eq.~(\ref{eq2-1-2}), $\alpha$ is the fine structure constant and $\bar{X}$ is the invariant transition probability of the Compton scattering, which will be discussed later in this section.  In this paper, we use the natural unit $\hbar=c=1$, unless otherwise stated explicitly.

  According to Ref.~\cite{noza09}, Eq.~(\ref{eq2-1-1}) is simplified in the Thomson approximation as follows:
\begin{eqnarray}
&&\hspace{-10mm}
\frac{{\rm{d}} n_{\rm{NK}}(x)}{{\rm{d}} \tau} = \frac{3}{64 \pi^{2}} \int {\rm{d}}^{3} p \, p_{e}(E) \, \frac{1}{\gamma^{2}} \int {\rm{d}} \Omega_{k^{\prime}} \bar{X}_{A}  \nonumber  \\
&&\hspace{20mm}
\times \frac{1-\beta \mu}{(1 - \beta \mu^{\prime})^{2}}  \left[ n(x^{\prime}) - n(x) \right]  \, ,
\label{eq2-1-3}  \\
&&\hspace{-9mm}
\frac{x^{\prime}}{x} \approx \frac{1 - \beta \mu}{1 - \beta \mu^{\prime}}  \, ,
\label{eq2-1-4}  \\
&&\hspace{-9mm}
{\rm{d}} \tau = N_{e} \sigma_{T} \, {\rm{d}} t  \, ,
\label{eq2-1-5}
\end{eqnarray}
where $x = \omega/k_{B}T_{\rm CMB}$, $x^{\prime} = \omega^{\prime}/k_{B} T_{\rm CMB}$, $\sigma_{T}=8\pi \alpha^{2}/(3m^{2})$ and $f(E)=N_{e} \pi^{2} p_{e}(E)$.  In Eq.~(\ref{eq2-1-3}), $\bvec{\beta} \equiv \bvec{p}/E$ is the velocity of the initial electron, $\gamma=1/\sqrt{1-\beta^{2}}$, $\mu = \bvec{\hat{\beta}} \cdot \bvec{\hat{k}}$ and $\mu^{\prime} = \bvec{\hat{\beta}} \cdot \bvec{\hat{k}}^{\prime}$.  The invariant transition probability $\bar{X}$ is reduced to
\begin{eqnarray}
\bar{X}_{A}  =  2 - \frac{2(1-{\rm cos}\Theta)}
{\gamma^2(1-\beta\mu)(1-\beta\mu^{\prime})}  \nonumber  \\
&&\hspace{-42mm}
+ \frac{(1-{\rm cos} \Theta)^2}{\gamma^4(1-\beta\mu)^2(1-\beta\mu^{\prime})^2} 
 \, ,
\label{eq2-1-6}
\end{eqnarray}
\begin{equation}
{\rm cos}\Theta = \mu\mu^{\prime}+\sqrt{1-\mu^2} \sqrt{1-\mu^{\prime 2}}\cos(\phi_{k}-\phi_{k^{\prime}})  \, .
\label{eq2-1-7}
\end{equation}
Note that $\bar{X}_{A}$ does not depend on $\omega$ and $\omega^{\prime}$.  Note also that the Thomson approximation used in deriving Eqs.~(\ref{eq2-1-3}) and (\ref{eq2-1-4}) is an extremely good approximation for the CMB photon energies.

  Applying the Fokker-Planck expansion to $n(x^{\prime})$, namely,
\begin{eqnarray}
n(x^{\prime}) = n(x) + \sum_{\ell=1}^{\infty} \, \frac{1}{\ell!} \, (x^{\prime}-x)^{\ell} \, n^{(\ell)}(x)  \, ,
\label{eq2-18}
\end{eqnarray}
one can rewrite Eq.~(\ref{eq2-1-3}) as follows:
\begin{eqnarray}
&&\hspace{-12mm}
\frac{{\rm{d}} n_{\rm{NK}}(x)}{{\rm{d}} \tau} = \frac{3}{64 \pi^{2}} \int {\rm{d}}^{3} p \, p_{e}(E) \, \frac{1}{\gamma^{2}} \int {\rm{d}} \Omega_{k^{\prime}} \bar{X}_{A} \, I_{\rm{NK}}(x)  \, ,
\label{eq2-1-9}  \\
&&\hspace{-12mm}
I_{\rm{NK}}(x) = \frac{1-\beta \mu}{(1 - \beta \mu^{\prime})^{2}} \sum_{\ell=1}^{\infty} \frac{1}{\ell!} \, \left(\beta \frac{\mu^{\prime}-\mu}{1-\beta \mu^{\prime}} \right)^{\ell} x^{\ell} \, n^{(\ell)}(x) \, ,
\label{eq2-1-10}
\end{eqnarray}
where $n^{(\ell)}(x)$ is the $n$-th order derivative of $n(x)$.

\subsection{Calculation in the electron rest frame}

  Calculation of Eq.~(\ref{eq2-1-9}) can be simplified by applying the following Lorentz transformations between angles in the CMB and electron rest frames \cite{noza09}:
\begin{eqnarray}
\mu = \frac{- \mu_{0} + \beta}{1 - \beta \mu_{0}}  \, ,
\label{eq2-2-1}  \\
\mu^{\prime} = \frac{- \mu^{\prime}_{0} + \beta}{1 - \beta \mu^{\prime}_{0}}  \, ,
\label{eq2-2-2}
\end{eqnarray}
which give
\begin{eqnarray}
&&\hspace{-83mm}
\bar{X}_{A} = 1 + \cos^{2} \Theta_{0}  \, ,
\label{eq2-2-3}  \\
\cos \Theta_{0} = \mu_{0} \mu^{\prime}_{0} + \sqrt{1-\mu^{2}_{0}} \sqrt{1-\mu^{\prime 2}_{0}} \cos (\phi_{k}-\phi_{k^{\prime}})  \, .
\label{eq2-2-4}
\end{eqnarray}
Inserting Eqs.~(\ref{eq2-2-1})--(\ref{eq2-2-4}) into Eq.~(\ref{eq2-1-9}), one has
\begin{eqnarray}
&&\hspace{-10mm}
\frac{{\rm{d}} n_{\rm{NK}}(x)}{{\rm{d}} \tau} = \int_{0}^{\infty} {\rm{d}} p \, p^{2} \, p_{e}(E) \, F_{\rm{NK}}(x)  \, ,
\label{eq2-2-5}  \\
&&\hspace{-10mm}
F_{\rm{NK}}(x) = \sum_{\ell=1}^{\infty} \, F_{\ell}(\beta) \, x^{\ell} \, n^{(\ell)}(x)  \, ,
\label{eq2-2-6}  \\
&&\hspace{-10mm}
F_{\ell}(\beta) = \int_{-1}^{+1} {\rm{d}} \mu_{0} \int_{-1}^{+1} {\rm{d}} \mu^{\prime}_{0} \, f_{\ell}(\beta, \mu_{0}, \mu^{\prime}_{0})  \, ,
\label{eq2-2-7}  \\
&&\hspace{-10mm}
f_{\ell}(\beta, \mu_{0}, \mu^{\prime}_{0}) = \frac{3}{16 \ell!} \frac{\beta^{\ell}}{\gamma^{4} (1 - \beta \mu_{0})^{\ell+3}} \, (\mu_{0} - \mu^{\prime}_{0})^{\ell}
  \nonumber  \\
&&\hspace{10mm}
\times [1 + \mu^{2}_{0} \mu^{\prime 2}_{0} + \frac{1}{2} (1 - \mu^{2}_{0})(1 - \mu^{\prime 2}_{0}) ]  \, .
\label{eq2-2-8}
\end{eqnarray}

  Equation (\ref{eq2-2-7}) can be integrated analytically.  We summarize the results up to $\mathcal{O}(\beta^{10})$ as follows:
\begin{eqnarray}
&&\hspace{-80mm}
F_{1}(\beta) = \frac{4}{3} \beta^{2} + \frac{4}{3} \beta^{4} + \frac{4}{3} \beta^{6} + \frac{4}{3} \beta^{8} + \frac{4}{3} \beta^{10}   \, ,
\label{eq2-2-9}  \\
&&\hspace{-80mm}
F_{2}(\beta) = \frac{1}{3} \beta^{2} + \frac{26}{15} \beta^{4} + \frac{47}{15} \beta^{6} + \frac{68}{15} \beta^{8} + \frac{89}{15} \beta^{10}  \, ,
\label{eq2-2-10}  \\
&&\hspace{-80mm}
F_{3}(\beta) = \frac{14}{25} \beta^{4} + \frac{172}{75} \beta^{6} + \frac{26}{5} \beta^{8} + \frac{232}{25} \beta^{10}  \, ,
\label{eq2-2-11}  \\
F_{4}(\beta) = \frac{7}{150} \beta^{4} + \frac{17}{25} \beta^{6} + \frac{1709}{630} \beta^{8} + \frac{10958}{1575} \beta^{10}  \, ,
\label{eq2-2-12}  \\
&&\hspace{-80mm}
F_{5}(\beta) = \frac{44}{525} \beta^{6} + \frac{7892}{11025} \beta^{8} + \frac{4472}{1575} \beta^{10}  \, ,
\label{eq2-2-13}  \\
&&\hspace{-80mm}
F_{6}(\beta) = \frac{11}{3150} \beta^{6} + \frac{239}{2450} \beta^{8} + \frac{117}{175} \beta^{10}  \, ,
\label{eq2-2-14}  \\
&&\hspace{-80mm}
F_{7}(\beta) = \frac{128}{19845} \beta^{8} + \frac{1832}{19845} \beta^{10} \, ,
\label{eq2-2-15}  \\
&&\hspace{-80mm}
F_{8}(\beta) = \frac{16}{99225} \beta^{8} + \frac{724}{99225} \beta^{10}   \, ,
\label{eq2-2-16}  \\
&&\hspace{-80mm}
F_{9}(\beta) = \frac{2}{6615} \beta^{10}   \, ,
\label{eq2-2-17}  \\
&&\hspace{-80mm}
F_{10}(\beta) = \frac{1}{198450} \beta^{10}  \, .
\label{eq2-2-18}
\end{eqnarray}
Note that Eq.~(\ref{eq2-2-6}) with Eqs.~(\ref{eq2-2-9})--(\ref{eq2-2-18}) is identical to Eq.~(37) with Eq.~(41)--(45) of Nozawa \& Kohyama \cite{noza14}.

\section{Delta function expansion formalism}

\subsection{Boltzmann equation}

In this section, we first review the prescription used in Brown \& Preston \cite{brow12}.  Let us start with the following identity relation:
\begin{eqnarray}
\frac{{\rm{d}}^{3}p^{\prime}}{2 E^{\prime}} = {\rm{d}}^{4} p^{\prime} \delta(p^{\prime 2} - m^{2})  \, .
\label{eq3-1-1}
\end{eqnarray}
Inserting Eq.~(\ref{eq3-1-1}) into Eq.~(\ref{eq2-1-1}) and eliminating the 4-dimensional delta function, one has
\begin{eqnarray}
&&\hspace{-12mm}
\frac{{\rm{d}} n(\omega)}{{\rm{d}} t} = - \int \frac{{\rm{d}}^{3}p}{(2\pi)^{3}E^{2}} \int {\rm{d}} \Omega_{k^{\prime}} \int {\rm{d}} \omega^{\prime} \, \frac{\omega^{\prime}}{\omega} \, \delta \Big( \omega^{\prime} - \omega  \nonumber  \\
&&\hspace{10mm}
+ \beta (\omega \mu - \omega^{\prime} \mu^{\prime}) + \frac{\omega \omega^{\prime}}{E} (1 - \cos \Theta) \Big)  \nonumber  \\
&&\hspace{0mm}
\times \, \alpha^{2} \, \bar{X} \left[ \, n(\omega) \left\{1 + n(\omega^{\prime}) \right\} f(E) \right.
 \nonumber  \\
&&\hspace{10mm}
\left. - \, n(\omega^{\prime}) \left\{1 + n(\omega) \right\} f(E^{\prime}) \right] \, .
\label{eq3-1-2}
\end{eqnarray}

  In the Thomson approximation $\gamma \omega/m \ll 1$, Eq.~(\ref{eq3-1-2}) is simplified as follows:
\begin{eqnarray}
&&\hspace{-12mm}
\frac{{\rm{d}} n_{\rm{BP}}(x)}{{\rm{d}} \tau} = \frac{3}{64 \pi^{2}} \int {\rm{d}}^{3} p \, p_{e}(E) \, \frac{1}{\gamma^{2}} \int {\rm{d}} \Omega_{k^{\prime}}  \bar{X}_{A}  \nonumber  \\
&&\hspace{4mm}
\times \int {\rm{d}}x^{\prime} \, \delta (x^{\prime} - x + \beta \, ( x \mu - x^{\prime} \mu^{\prime}) ) \, G(x^{\prime})  \, ,
\label{eq3-1-3}  \\
&&\hspace{-12mm}
G(x^{\prime}) = \frac{x^{\prime}}{x} \left[ n(x^{\prime}) - n(x) \right]  \, .
\label{eq3-1-4}
\end{eqnarray}
In Eq.~(\ref{eq3-1-3}), the delta function is formally expanded by
\begin{eqnarray}
&&\hspace{-12mm}
\delta (x^{\prime} - x + \mathcal{O}(x^{\prime})) = \sum_{\ell=0}^{\infty} \frac{1}{\ell!} \, \mathcal{O}^{\ell}(x^{\prime}) \, \delta^{(\ell)}(x^{\prime} - x)  \, ,
\label{eq3-1-5}  \\
&&\hspace{-12mm}
\mathcal{O}(x^{\prime}) = \beta \, (x \mu - x^{\prime} \mu^{\prime})  \, ,
\label{eq3-1-6}
\end{eqnarray}
where $\delta^{(\ell)} (x^{\prime}-x)$ is the $\ell$-th order derivative of $\delta(x^{\prime}-x)$ in terms of $x^{\prime}$.

  Inserting Eq.~(\ref{eq3-1-5}) into Eq.~(\ref{eq3-1-3}), one finally has
\begin{eqnarray}
&&\hspace{-13mm}
\frac{{\rm{d}} n_{\rm{BP}}(x)}{{\rm{d}} \tau} = \frac{3}{64 \pi^{2}} \int {\rm{d}}^{3} p \, p_{e}(E) \, \frac{1}{\gamma^{2}} \int {\rm{d}} \Omega_{k^{\prime}} \, \bar{X}_{A} \, I_{\rm{BP}}(x) \, ,
\label{eq3-1-7}  \\
&&\hspace{-12mm}
I_{\rm{BP}}(x) = \sum_{\ell=0}^{\infty} \frac{1}{\ell!} \int {\rm{d}}x^{\prime} \, \delta^{(\ell)} (x^{\prime} - x) \, G(x^{\prime}) \, \mathcal{O}^{\ell}(x^{\prime}) \, .
\label{eq3-1-8}
\end{eqnarray}
In Ref.~\cite{brow12}, Eq.~(\ref{eq3-1-2}) was calculated with the delta function identities defined in their Appendix C up to fourth-order derivatives.

  Thus, we have shown the expressions for the Boltzmann equation under the Thomson approximation in two different schemes.  Although the expressions for $I_{\rm{NK}}(x)$ and $I_{\rm{BP}}(x)$ are quite different each other, we will show in the next subsection that these formalisms are indeed identical.

\subsection{Equivalence of the formalisms}

  In order to show the equivalence of two formalisms, we start with $I_{\rm{BP}}(x)$ of Eq.~(\ref{eq3-1-8}) and derive the same expression as $I_{\rm{NK}}(x)$ of Eq.~(\ref{eq2-1-10}).  Let us first introduce the definition for the $\ell$-th order derivative of the delta function:
\begin{eqnarray}
&&\hspace{-12mm}
\int {\rm{d}} x \, \delta^{(\ell)}(x-a) \, \phi(x) = (-1)^{\ell} \int {\rm{d}} x\, \delta(x-a) \, \phi^{(\ell)}(x)  \,  ,
\label{eq3-2-1}
\end{eqnarray}
where $\phi(x)$ is an arbitrary function.  Then, one can rewrite Eq.~(\ref{eq3-1-8}) as follows:
\begin{eqnarray}
&&\hspace{-14mm}
I_{\rm{BP}}(x) = \sum_{\ell=0}^{\infty} \frac{(-1)^{\ell}}{\ell!} \int {\rm{d}} x^{\prime} \, \delta (x^{\prime}-x) \left[ G(x^{\prime}) \, \mathcal{O}^{\ell}(x^{\prime}) \right]^{(\ell)}  .
\label{eq3-2-2}
\end{eqnarray}
Equation (\ref{eq3-2-2}) is further simplified with the relation:
\begin{eqnarray}
&&\hspace{-12mm}
\left[ G(x^{\prime}) \,  \mathcal{O}^{\ell}(x^{\prime}) \right]^{(\ell)} = \sum_{j=0}^{\ell} \, _{\ell} {\rm{C}}_{j} \, \, G^{(j)}(x^{\prime})  \nonumber  \\
&&\hspace{+23mm}
\times \, \frac{\ell!}{j!} \, (-\beta \mu^{\prime})^{\ell-j} \, \mathcal{O}^{j}(x^{\prime})  \, .
\label{eq3-2-3}
\end{eqnarray}
Inserting Eq.~(\ref{eq3-2-3}) into Eq.~(\ref{eq3-2-2}), exchanging the order of summations for $j$ and $\ell$, and introducing a new index $i$ by $\ell=j+i$, one obtains
\begin{eqnarray}
&&\hspace{-12mm}
I_{\rm{BP}}(x) = \sum_{j=0}^{\infty} \, \frac{(-1)^{j}}{j!} \, \sum_{i=0}^{\infty} \, _{j+i} {\rm{C}}_{j} \, \, (\beta \mu^{\prime})^{i}  \nonumber  \\
&&\hspace{21mm}
\times \, \int {\rm{d}} x^{\prime} \, \delta (x^{\prime}-x) \, G^{(j)}(x^{\prime}) \, \mathcal{O}^{j} (x^{\prime})  \nonumber  \\
&&\hspace{-9mm}
= \sum_{j=0}^{\infty} \frac{(-1)^{j}}{j!} \frac{\mathcal{O}^{j}(x)}{(1-\beta \mu^{\prime})^{j+1}} \int {\rm{d}} x^{\prime} \delta (x^{\prime}-x) \, G^{(j)}(x^{\prime})  \, .
\label{eq3-2-4}
\end{eqnarray}
In deriving the last equality of Eq.~(\ref{eq3-2-4}), we used
\begin{eqnarray}
\sum_{i=0}^{\infty} \, _{j+i} {\rm{C}}_{j} \, \, (\beta \mu^{\prime})^{i} \, = \frac{1}{(1 - \beta \mu^{\prime})^{j+1}}  \, ,
\label{eq3-2-5}
\end{eqnarray}
and we took $\mathcal{O}^{j}(x)$ outside of the integral with a familiar property of the delta function.

  The $j$-th order derivative of $G(x^{\prime})$ is calculated with Eq.~(\ref{eq3-1-4}) as follows:
\begin{eqnarray}
G^{(j)}(x^{\prime}) = \frac{x^{\prime}}{x} \, n^{(j)}(x^{\prime}) + \frac{j}{x} \, n^{(j-1)}(x^{\prime})  \, ,
\label{eq3-2-6}
\end{eqnarray}
where $n^{(0)}(x^{\prime}) \equiv G(x^{\prime})$.  Inserting Eqs.~(\ref{eq3-1-6}) and (\ref{eq3-2-6}) into Eq.~(\ref{eq3-2-4}), one finally obtains
\begin{eqnarray}
&&\hspace{-13mm}
I_{\rm{BP}}(x) = \frac{1-\beta \mu}{(1-\beta \mu^{\prime})^{2}} \, \sum_{j=1}^{\infty} \frac{1}{j!} \left( \beta \frac{\mu^{\prime}-\mu}{1-\beta \mu^{\prime}} \right)^{j} x^{j} \, n^{(j)}(x)  \, ,
\label{eq3-2-7}
\end{eqnarray}
where $j=0$ term was zero because of the relation:
\begin{eqnarray}
\int {\rm{d}}x^{\prime} \delta(x^{\prime}-x) \, G(x^{\prime}) = 0  \, .
\label{eq3-2-8}
\end{eqnarray}
Thus, one finds that two formalisms are equivalent in the Thomson approximation

  Before closing this section it should be noted as follows:  In Nagirner \& Poutanen \cite{nagi94}, a similar discussion was made between the delta function expansion formalism and the Fokker-Planck expansion formalism, where the Maxwellian electron distribution function was expanded in terms $\theta_{e} \equiv k_{B}T_{e}/mc^{2}$ and the delta function, and the $O(\theta_{e})$ terms were retained.  On the other hand, the equivalence of the two formalisms in this paper is more general, which is independent of the electron distribution functions.

\section{Conclusion}

  We studied the SZ effect in the Thomson approximation.  In Section 2, we investigated the Boltzmann equation for the photon distribution function in the Fokker-Planck expansion approximation.  We derived the analytic formula for the Boltzmann equation which is identical to that obtained in Nozawa \& Kohyama \cite{noza14}.

  In Section 3, we studied the Boltzmann equation for the photon distribution function in the delta function expansion approximation \cite{brow12}.  Starting from the expression in the delta function expansion approximation, we derived the same expression as the Fokker-Planck expansion approximation.  We conclude that two expansion approximation formalisms are equivalent under the Thomson approximation, which is extremely good approximation for the CMB photon energies.  This will clarify the situation for existing theoretical methods to analyse observation data.

\section*{Acknowledgements}
We thank our referee for a valuable suggestion.

\bibliographystyle{elsarticle-num}

\begin{thebibliography}{00}
\bibitem{zeld69} Ya. B. Zeldovich, R. A. Sunyaev, Ap\&SS 4 (1969) 301.
\bibitem{suny80} R. A. Sunyaev, Ya. B. Zeldovich, MNRAS 190 (1980) 413.
\bibitem{birk99} M. Birkinshaw, Phys. Rep. 310 (1999) 97.
\bibitem{carl02} J. E. Carlstrom, G. P. Holder, E. D. Reese, ARA\&A 40 (2002) 643.
\bibitem{komp56} A. S. Kompaneets, Sov. Phys.$-$JETP 31 (1956) 876.
\bibitem{alle02} S. W. Allen, R. W. Schmidt, A. C. Fabian, MNRAS 335 (2002) 256
\bibitem{wrig79} E. L. Wright, ApJ 232 (1979) 348.
\bibitem{reph95} Y. Rephaeli, ApJ 445 (1995) 33.
\bibitem{chal98} A. Challinor, A. Lasenby, ApJ 499 (1998) 1.
\bibitem{itoh98} N. Itoh, Y. Kohyama, S. Nozawa, ApJ 502 (1998) 7.
\bibitem{noza98} S. Nozawa, N. Itoh, Y. Kohyama, ApJ 508 (1998) 17.
\bibitem{itoh00} N. Itoh, S. Nozawa, Y. Kohyama, ApJ 533 (2000) 588.
\bibitem{suny81} R. A. Sunyaev, Ya. B. Zeldovich, Astrophys. Space Phys. Rev. 1 (1981) 1.
\bibitem{chlu05} J. Chluba, G. H\"utsi, R. A. Sunyaev, A\&A 434 (2005) 811.
\bibitem{noza05} S. Nozawa, N. Itoh, Y. Kohyama, A\&A 440 (2005) 39.
\bibitem{brow12} L. S. Brown, D. L. Preston, Astroparticle Phys. 35 (2012) 742.
\bibitem{noza09} S. Nozawa, Y. Kohyama, Phys. Rev. D 79 (2009) 083005.
\bibitem{noza14} S. Nozawa, Y. Kohyama, MNRAS 441 (2014) 3018.
\bibitem{nagi94} D. I. Nagirner, J. Poutanen, Astrophys. Space Phys. Rev. 9 (1994) 1.

\end{thebibliography}

\end{document}